\documentclass[12pt]{article}
\usepackage{graphicx,amssymb,amsfonts,amsmath,epsfig,color}

\makeatletter
\DeclareSymbolFont{AMSb}{U}{msb}{m}{n}
\DeclareSymbolFontAlphabet{\mathbb}{AMSb}
\renewcommand{\section}{\@startsection{section}{1}{\z@}%
                                    {-7ex \@plus -1ex \@minus -.2ex}%
                                    {2.5ex \@plus.2ex}%
                                    {\normalfont\large\scshape\centering}}
\renewcommand{\subsection}{\@startsection{subsection}{2}{\z@}%
                                       {-5ex \@plus -1ex \@minus -.2ex}%
                                       {1.5ex \@plus.2ex}%
                                       {\normalfont\normalsize\scshape}}
\renewcommand{\subsubsection}{\@startsection{subsubsection}{3}{\z@}%
                                       {-5ex \@plus -1ex \@minus -.2ex}%
                                       {1.5ex \@plus.2ex}%
                                       {\normalfont\normalsize\scshape}}

\renewcommand\@seccntformat[1]{\ignorespaces\csname #1name\endcsname\space
                               \csname the#1\endcsname.\quad}   
\newdimen\captionmargin
\newdimen\captionindent
\newdimen\captionwidth
\newcommand{\captionfont}{\slshape}
\newcommand\@captionlabel[1]{\textsc{#1:}\space}
\long\def\@makecaption#1#2{%
  \vskip\abovecaptionskip
  \captionwidth\hsize
  \advance\captionwidth -2\captionmargin
  \sbox\@tempboxa{\@captionlabel{#1}\captionfont #2}%
  \ifdim \wd\@tempboxa >\captionwidth
    \ifdim\captionindent>\z@
      \advance\captionwidth -\captionindent
      \hskip\captionindent
    \fi
    \hskip\captionmargin
    \parbox[t]{\captionwidth}{\leavevmode\hskip-\captionindent
      \@captionlabel{#1}\captionfont #2}%
  \else
    \global \@minipagefalse
    \hb@xt@\hsize{\hfil\box\@tempboxa\hfil}%
  \fi
  \vskip\belowcaptionskip}
\def\eqnarray{%
   \stepcounter{equation}%
   \def\@currentlabel{\p@equation\theequation}%
   \global\@eqnswtrue
   \m@th
   \global\@eqcnt\z@
   \tabskip\@centering
   \let\\\@eqncr
   $$\everycr{}\halign to\displaywidth\bgroup
       \hskip\@centering$\displaystyle\tabskip\z@skip{##}$\@eqnsel
      &\global\@eqcnt\@ne$\;\hfil{##}$\hfil
      &\global\@eqcnt\tw@$\;\displaystyle{##}$\hfil\tabskip\@centering
      &\global\@eqcnt\thr@@ \hb@xt@\z@\bgroup\hss##\egroup
         \tabskip\z@skip
      \cr}
\ifcase \@ptsize
\or
\or
\fi
  \divide\@tempdima\baselineskip
  \@tempcnta=\@tempdima
\makeatother
\begin{document}

%
%

\renewcommand{\theequation}{\arabic{section}.\arabic{equation}}
\renewcommand{\thefigure}{\arabic{figure}}
\newcommand{\gapprox}{%
\mathrel{%
\setbox0=\hbox{$>$}\raise0.6ex\copy0\kern-\wd0\lower0.65ex\hbox{$\sim$}}}
\textwidth 165mm \textheight 220mm \topmargin 0pt \oddsidemargin 2mm
\def\ib{{\bar \imath}}
\def\jb{{\bar \jmath}}

\newcommand{\ft}[2]{{\textstyle\frac{#1}{#2}}}
\newcommand{\be}{\begin{equation}}
\newcommand{\ee}{\end{equation}}
\newcommand{\bea}{\begin{eqnarray}}
\newcommand{\eea}{\end{eqnarray}}
\newcommand{\Identity}{{1\!\rm l}}
\newcommand{\cx}{\overset{\circ}{x}_2}
\def\CN{$\mathcal{N}$}
\def\CH{$\mathcal{H}$}
\def\hg{\hat{g}}
\newcommand{\bref}[1]{(\ref{#1})}
\def\espai{\;\;\;\;\;\;}
\def\zespai{\;\;\;\;}
\def\avall{\vspace{0.5cm}}
\newtheorem{theorem}{Theorem}
\newtheorem{acknowledgement}{Acknowledgment}
\newtheorem{algorithm}{Algorithm}
\newtheorem{axiom}{Axiom}
\newtheorem{case}{Case}
\newtheorem{claim}{Claim}
\newtheorem{conclusion}{Conclusion}
\newtheorem{condition}{Condition}
\newtheorem{conjecture}{Conjecture}
\newtheorem{corollary}{Corollary}
\newtheorem{criterion}{Criterion}
\newtheorem{defi}{Definition}
\newtheorem{example}{Example}
\newtheorem{exercise}{Exercise}
\newtheorem{lemma}{Lemma}
\newtheorem{notation}{Notation}
\newtheorem{problem}{Problem}
\newtheorem{prop}{Proposition}
\newtheorem{rem}{{\it Remark}}
\newtheorem{solution}{Solution}
\newtheorem{summary}{Summary}
\numberwithin{equation}{section}
\newenvironment{pf}[1][Proof]{\noindent{\it {#1.}} }{\ \rule{0.5em}{0.5em}}
\newenvironment{ex}[1][Example]{\noindent{\it {#1.}}}

\thispagestyle{empty}


\begin{center}

{\LARGE\scshape On regular rotating black holes\par}
\vskip15mm

\textsc{R. Torres and F. Fayos}
\par\bigskip
{\em
Department of Physics, UPC, Barcelona, Spain.}\\[.1cm]
\vspace{5mm}

\end{center}

\section*{Abstract}
Different proposals for regular rotating black hole spacetimes have appeared recently in the literature. However, a rigorous analysis and proof of the regularity of this kind of spacetimes is still lacking. In this note we analyze rotating Kerr-like black hole spacetimes and find the necessary and sufficient conditions for the regularity of all their second order scalar invariants polynomial in the Riemann tensor. We also show that the regularity is linked to a violation of the weak energy conditions around the core of the rotating black hole.

\vskip10mm
\noindent KEYWORDS: Black holes, regular black holes, singularities, invariants, energy conditions.

\vspace{3mm} \vfill{ \hrule width 5.cm \vskip 2.mm {\small
\noindent E-mail: ramon.torres-herrera@upc.edu, f.fayos@upc.edu}}

\newpage
\setcounter{page}{1}

\setcounter{equation}{0}

\section{Introduction}

Most astrophysically significant
bodies are rotating. If a rotating body collapses, the rate of rotation
will speed up, maintaining constant angular momentum. If the body finally generates a black hole it will be a rotating black hole. From a classical point of view (\textit{no-hair conjecture}) the resulting spacetime will be described by a Kerr or a Kerr-Newman solution (in the charged case) [see, for instance, \cite{Griff}]. This implies the existence of certain horizons, a specific causal structure and a \emph{singular ring}.


Several authors have suggested that the existence of singularities in the classical solutions has to be considered as a weakness of the theory rather than as a real physical prediction.
Some authors have tried to solve this by introducing non-standard energy-momentum tensors mainly acting in the core of the black hole (see, for example, \cite{Bardeen}\cite{A-BI}\cite{A-BII}\cite{B&V}). However, most authors expect that the inclusion of quantum theory in the description of black holes could avoid the existence of their singularities (see, for example, \cite{B&R}\cite{A&B2005}\cite{Hay2006}\cite{Frolov2014}\cite{dust2014}\cite{G&P2014}\cite{H&R2014} and references therein).

In this way, recently there have appeared different proposals
for \emph{regular} rotating black holes spacetimes (see, for instance, \cite{R&T}\cite{B&M}\cite{A-A}\cite{Tosh}\cite{D&G}\cite{Ghosh}\cite{LGS}). In order to check the kindness of these proposals one should check for the absence of \textit{scalar curvature singularities}. We say that there is a scalar curvature singularity
in the spacetime if any scalar invariant polynomial in the Riemann tensor diverges when approaching it along any incomplete curve. In the literature one finds that the authors use to check one, two or even three \textit{standard} invariants\footnote{Unfortunately, when this is the case, it is not unusual that the authors check invariants that are algebraically dependent, so that the extra-check is redundant.} along one or two directions approaching the possible singularity. Indeed, this is a \textit{necessary} condition that a regular black hole should satisfy. However, this is not a \emph{sufficient} condition to guarantee the absence of scalar curvature singularities.
It is well-known \cite{Weinberg} that an arbitrary spacetime possesses at most 14 second order algebraically independent invariants. The finiteness of \emph{all} the invariants is a necessary and sufficient condition for the absence of scalar curvature singularities.

Our aim in this note is to rigourously prove the conditions for the absence of scalar curvature singularities in rotating Kerr-like black hole spacetimes. This will require the study of some of their properties, what will provide us with a complete set of algebraically independent curvature scalars for these spacetimes. Then, we will try to provide a necessary and sufficient condition for the absence of scalar curvature singularities when the metric is written in the usual Boyer-Lindquist-like coordinates. Finally, since most of the conditions of the standard singularity theorems can be satisfied by regular rotating black holes (usually, the causality conditions and the boundary/initial conditions), it seems appropriate to look for the non-fulfillment of energy conditions as the culprit for the absence of singularities. With this goal, we will analyze the violation of the weak energy conditions in these spacetimes.

The note has been divided as follows. In section \ref{secRBH} we will define the general class of stationary rotating black hole spacetimes that will be the subject of the work and we will analyze some of their properties. In section \ref{secSCS} we will find the algebraically independent second order invariants of the spacetime and, by using them, we will introduce a necessary and sufficient condition for the absence of scalar curvature singularities. In section \ref{secWEC} we will analyze the relationship between the absence of singularities and the violations of the weak energy conditions in the core of rotating black holes. Finally, the conclusions are collected in section \ref{secCon}.

\section{Kerr-like rotating black hole spacetimes}\label{secRBH}

Different metrics corresponding to stationary rotating black holes (RBH) have been proposed. While they have been obtained by different approaches, most of them share a common Kerr-like form (see, for instance, \cite{R&T}\cite{B&M}\cite{A-A}\cite{Tosh}\cite{D&G}\cite{Ghosh}\cite{LGS}). The general metric corresponding to this kind of RBH takes the form, in Boyer-Lidquist-like coordinates,
\begin{equation}\label{gIKerr}
ds^2=-\frac{\Delta_r}{\Sigma} (dt-a \sin^2\theta d\phi)^2+
\frac{\Sigma}{\Delta_r} dr^2+\Sigma d\theta^2+\frac{\sin^2\theta}{\Sigma}(a dt-(r^2+a^2)d\phi)^2,
\end{equation}
where
\[
\Sigma=r^2+a^2 \cos^2\theta, \hspace{1cm} \Delta_r=r^2-2 \mathcal M(r) r+a^2,
\]
$\mathcal M(r)$ is assumed to be a $C^3$ function and $a$ ($\neq 0$) is a constant (which is interpreted as a \textit{rotation parameter} by using the same arguments as for the Kerr solution). Note that this metric reduces to the Kerr solution if $\mathcal M(r)=m$=constant and that it reduces to the (charged) Kerr-Newman solution if $\mathcal M(r)=m-e^2/(2r)$.

Different articles dealing with regular rotating black holes propose different forms for the function $\mathcal M(r)$. Its exact expression depends on the procedure used to obtain the RBH. In many cases the authors just propose heuristic forms for $\mathcal M$ in order to try to avoid the existence of singularities (for instance, \cite{B&M}\cite{A-A}\cite{LGS}), in other cases a physical approach provides a specific $\mathcal M(r)$. Among these physical approaches there are two main paths to $\mathcal M(r)$. For some authors (for instance, \cite{Tosh}\cite{D&G}\cite{Ghosh}) $\mathcal M(r)=G_0 m(r)$, where we have now made explicit Newton's gravitational constant $G_0$ and $m(r)$ is a \textit{mass-energy} function\footnote{\textit{Mass-energy} and not just the mass, as the Kerr-Newman case exemplifies.} provided by the specific approach. For others (see, for instance, \cite{R&T}) $\mathcal M(r)=G(r) m$, where $m$ is the (constant) black hole mass and $G(r)$ is a \textit{running} Newton constant (i.e., Newton's constant is replaced by an appropriate scale dependent function).

In order to analyze the general properties of the RBH spacetime it will be convenient to describe the spacetime in a Newman-Penrose formalism. For this goal we first provide the following null tetrad-frame:
\begin{eqnarray*}
\mathbf{l} &=&\frac{1}{\Delta_r} \left( (r^2+a^2) \frac{\partial}{\partial t}+\Delta_r \frac{\partial}{\partial r}+a \frac{\partial}{\partial \phi}\right),\\
\mathbf k &=&\frac{1}{2 \rho^2} \left( (r^2+a^2) \frac{\partial}{\partial t}-\Delta_r \frac{\partial}{\partial r}+a \frac{\partial}{\partial \phi}\right),\\
\mathbf m &=& \frac{1}{ \sqrt{2} \varrho } \left(i a \sin\theta \frac{\partial}{\partial t}+\frac{\partial}{\partial \theta}+i \csc\theta \frac{\partial}{\partial \phi} \right),\\
\mathbf{\bar m} &=& \frac{1}{ \sqrt{2} \bar\varrho } \left(-i a \sin\theta \frac{\partial}{\partial t}+\frac{\partial}{\partial \theta}-i \csc\theta \frac{\partial}{\partial \phi} \right),
\end{eqnarray*}
where $\varrho\equiv r+i a \cos\theta$, $\bar\varrho\equiv r-i a \cos\theta$ and the tetrad is normalized as follows $\mathbf l^2=\mathbf k^2=\mathbf m^2=\mathbf{\bar m}^2=0$ and $\mathbf l\cdot \mathbf k=-1= -\mathbf{m}\cdot \mathbf{\bar m}$.

\begin{prop}\label{PTD}
The RBH metric (\ref{gIKerr}) is Petrov type D and the two double principal null directions are $\mathbf l$ and $\mathbf k$.
\end{prop}

In order to show this, we define in the usual way the five complex scalar fuctions determining the Weyl tensor ($C_{\kappa\lambda\mu\nu}$):
\begin{eqnarray*}
\Psi_0&=&C_{\kappa\lambda\mu\nu} l^\kappa m^\lambda l^\mu m^\nu,\\
\Psi_1&=&C_{\kappa\lambda\mu\nu} l^\kappa k^\lambda l^\mu m^\nu,\\
\Psi_2&=&C_{\kappa\lambda\mu\nu} l^\kappa m^\lambda \bar{m}^\mu k^\nu,\\
\Psi_3&=&C_{\kappa\lambda\mu\nu} k^\kappa l^\lambda k^\mu \bar{m}^\nu,\\
\Psi_4&=&C_{\kappa\lambda\mu\nu} k^\kappa \bar{m}^\lambda k^\mu \bar{m}^\nu.
\end{eqnarray*}

Now it its easy to check that we have the following relations for the RBH metric (\ref{gIKerr})
\begin{equation}\label{Psi2}
\Psi_2=-\frac{6 \varrho \mathcal M  - 2 (r+\varrho) \bar{\varrho} \mathcal M' +r \bar{\varrho}^{2}\mathcal M'' }{6 \varrho\bar{\varrho}^{3}}.
\end{equation}
and (checking the two double principal null directions)
\begin{eqnarray*}
\Psi_0=\Psi_1=0, (\Psi_2\neq 0) \Leftrightarrow C_{\alpha\beta\gamma[\delta} k_{\mu]} k^\beta k^\gamma =0,\\
\Psi_4=\Psi_3=0, (\Psi_2\neq 0) \Leftrightarrow C_{\alpha\beta\gamma[\delta} l_{\mu]} l^\beta l^\gamma =0.
\end{eqnarray*}
The relations show that proposition \ref{PTD} is indeed satisfied \cite{Kramer}. $\hspace{1cm}\Box$

\begin{prop}
The metric (\ref{gIKerr}) with $\mathcal M\neq$constant is Segre type [(1,1) (1 1)].
\end{prop}

 Note that the $\mathcal M\neq$constant case is precisely the case we are interested in, since the $\mathcal M=$constant (i.e., Kerr's case) is known to be singular.

Now, the ten independent components of the Ricci tensor can be determined by the curvature scalar $\mathcal{R}$, the scalars
\begin{eqnarray*}
\Phi_{00}&=&\frac{1}{2} R_{\mu\nu} l^\mu l^\nu,\hspace{1cm} \Phi_{22}=\frac{1}{2} R_{\mu\nu} k^\mu k^\nu\\
\Phi_{01}&=&\frac{1}{2} R_{\mu\nu} l^\mu m^\nu,\hspace{1cm} \Phi_{12}=\frac{1}{2} R_{\mu\nu} k^\mu m^\nu\\
\Phi_{02}&=&\frac{1}{2} R_{\mu\nu} m^\mu m^\nu,\hspace{1cm} \Phi_{11}=\frac{1}{4} R_{\mu\nu} (k^\mu l^\nu+m^\mu \bar{m}^\nu),
\end{eqnarray*}
and the relationship $\bar{\Phi}_{AB}=\Phi_{BA}$.

In our case it is easy to check that $\Phi_{00}=\Phi_{12}=\Phi_{01}=\Phi_{22}=\Phi_{02}=0$ and
\begin{equation}\label{Phi11}
\Phi_{11}=\frac{2 (r^2-a^2 \cos^2\theta) \mathcal M' - r \Sigma \mathcal M''}{4 \Sigma^2}.
\end{equation}
This already shows the proposition \cite{Kramer}\cite{MFL}. Nevertheless, let us explicit for later use that
we can define a real orthonormal basis $\{\mathbf{t}, \mathbf{x}, \mathbf{y}, \mathbf{z } \}$
formed by a timelike vector $\mathbf{t}\equiv (\mathbf{l}+\mathbf{k})/\sqrt{2}$ and three spacelike vectors: $\mathbf{z}\equiv (\mathbf{l}-\mathbf{k})/\sqrt{2}$,  $\mathbf x=(\mathbf m +\bar{\mathbf m})/\sqrt{2}$ and $\mathbf y=(\mathbf m -\bar{\mathbf m}) i/\sqrt{2}$. Then,
$\mathbf t$ and $\mathbf z$ are two eigenvectors of the Ricci tensor with eigenvalue
\begin{equation}\label{lambda1}
\lambda_1=\frac{2 a^2 \cos^2{\theta} \mathcal M'+r \Sigma \mathcal M''}{\Sigma^2}.
\end{equation}
$\mathbf x$ and $\mathbf y$ are two eigenvectors of the Ricci tensor with eigenvalue
\begin{equation}\label{lambda2}
\lambda_2=\frac{2 r^2 \mathcal M'}{\Sigma^2}
\end{equation}
and the two different eigenvalues are related to $\Phi_{11}$ through
\[
\Phi_{11}=-\frac{1}{4} (\lambda_1-\lambda_2).
\]
In this way, the Ricci tensor can be written as
\begin{equation}\label{Ricci}
R_{\mu\nu}= \lambda_1\, (-t_\mu t_\nu+z_\mu z_\nu)+ \lambda_2 (x_\mu x_\nu+y_\mu y_\nu). \hspace{1cm} \Box
\end{equation}

With regard to the question of the regularity of the spacetime, one immediately sees that the \emph{metric} (\ref{gIKerr}) is singular if there are values of $r$ such that $\Delta_r=0$ and if $\Sigma=0$. On the one hand, $\Delta_r=0$ is not a curvature singularity since, as we will see, the curvature scalars do not diverge for the values $r$ ($\neq 0$) where $\Delta_r=0$. Nevertheless, the coordinate $r$ changes its character from spacelike when $\Delta_r>0$ to timelike when $\Delta_r<0$. Therefore, the boundaries $\Delta_r=0$ between these regions are simply horizons.

On the other hand, curvature singularities do may appear if  $\Sigma=0$ or, in other words, in $(r=0,\theta=\pi/2)$.This is confirmed by the scalars that we have been computing so far [(\ref{Psi2}),(\ref{Phi11}),(\ref{lambda1}),(\ref{lambda2})]. Note that, in case there were a singularity at $(r=0,\theta=\pi/2)$, it would be, as in the Kerr case, a \textit{ring singularity}. Therefore, independently of whether the spacetime is singular or not, $r=0$ can be crossed (through $\theta\neq \pi/2$) and we can analytically extend the metric beyond $r=0$ by just considering negative values for the coordinate $r$.

\section{Avoiding scalar curvature singularities}\label{secSCS}

As stated in the introduction, an arbitrary spacetime possesses at most 14 second order algebraically independent invariants. From a historical point of view, many authors have proposed different sets of invariants as algebraically independent. In most cases it has been later shown that this was not the case. Fortunately, a minimum set of reliable independent invariants for the RBH spacetime exists. This can be shown thanks to the following result by Zakhary and McIntosh \cite{ZM}


\begin{prop}
The algebraically complete set of second order invariants for a Petrov type D spacetime and Segre type [(1,1) (1 1)] is $\{\mathcal R,I,I_6,K\}$.\hspace{2cm} $\Box$
\end{prop}
Apart form the already defined curvature scalar $\mathcal R$, the rest of the invariants are defined as\footnote{Here the invariants are written in tensorial form. See \cite{ZM} for their spinorial form.}
\begin{eqnarray*}
I_6&\equiv&
\frac{1}{12}
{S_\alpha}^\beta {S_\beta}^\alpha,\\
I &\equiv&
\frac{1}{24}
\bar{C}_{\alpha\beta\gamma\delta}
\bar{C}^{\alpha\beta\gamma\delta},\\
K &\equiv&
\frac{1}{4}
\bar{C}_{\alpha\gamma\delta\beta}
S^{\gamma\delta} S^{\alpha\beta},
\end{eqnarray*}
where ${S_\alpha}^\beta \equiv {R_\alpha}^\beta-
{\delta_\alpha}^\beta \mathcal{R}/4$ and
$\bar{C}_{\alpha\beta\gamma\delta}\equiv
(C_{\alpha\beta\gamma\delta} +i\ *C_{\alpha\beta\gamma\delta})/2$
is the complex conjugate of the selfdual Weyl tensor
being $*C_{\alpha\beta\gamma\delta}\equiv
\epsilon_{\alpha\beta\mu\nu} C^{\mu\nu}_{\ \ \gamma\delta}/2$ the dual of the Weyl tensor.
Note that $\mathcal R$ and $I_6$ are real, while $I$ and $K$ are complex. Therefore for this type of spacetimes there are only 6 independent real scalars.

It trivially follows from our previous propositions
\begin{corollary}
The algebraically complete set of second order invariants for the RBH metric (\ref{gIKerr}) is $\{\mathcal R,I,I_6,K\}$.
\end{corollary}

Now we can use this result to get a necessary and sufficient condition for the absence of scalar curvature singularities:

\begin{theorem}\label{teorema}
Assuming a RBH metric (\ref{gIKerr}) possessing a $C^3$ function $\mathcal M(r)$, all its second order curvature invariants will be finite at $(r=0,\theta=\pi/2)$ if, and only if,
\begin{equation}\label{condisreg}
 \mathcal M (0)= \mathcal M' (0)= \mathcal M'' (0)=0 .
\end{equation}
\end{theorem}

First, note that, even if there is a singularity at $(r=0,\theta=\pi/2)$, we can define $\mathcal M$, $\mathcal M'$ and $\mathcal M''$ at $r=0$ since $(r=0,\theta\neq\pi/2)$ belongs to the spacetime.
The theorem can be shown by just analyzing the set of scalars for the particular case. In this way, by computing the curvature scalar for a RBH one finds
\[
\mathcal{R}=\frac{2 (2 \mathcal M' + r \mathcal M'')}{\Sigma}.
\]
For this to be finite along a path approaching $(r=0,\theta=\pi/2)$ it is necessary that $\mathcal M'(0)=\mathcal M''(0)=0$. Let us now define the dimensionless quantity $\xi\equiv a \cos\theta/r$ and $\xi^*$ its value in the limit along a chosen path approaching $r=0$. Now, assuming that the found necessary conditions are satisfied, we find
\begin{eqnarray*}
\mathcal R &\rightarrow& \frac{4 \mathcal M'''(0)}{1+\xi^{*2}} \hspace{1cm} \mbox{ if } \xi^* \mbox{ finite and }\\
\mathcal R &\rightarrow& 0  \hspace{2cm} \mbox{ if } \xi^* \mbox{ infinite}.
\end{eqnarray*}
As can be seen, $\mathcal R$ would be finite along any path.

On the other hand, if $\mathcal M'(0)=\mathcal M''(0)=0$ is satisfied then
\begin{eqnarray*}
I_6 &\rightarrow& \frac{\{\mathcal M'''(0)\}^2 \xi^{*4}}{3 (1+\xi^{*2})^4} \hspace{1cm}\mbox{ if } \xi^* \mbox{ finite and }\\
I_6 &\rightarrow& 0 \hspace{3cm} \mbox{ if } \xi^* \mbox{ infinite},
\end{eqnarray*}
what again is finite along any path.

The complex scalar $I$ takes the form
\[
I=\frac{[6 \varrho \mathcal M-2 (r+\varrho)\bar\varrho \mathcal M'+r \bar\varrho^{2} \mathcal M'']^2}{36 \bar\varrho^{6} \varrho^2 }.
\]
This adds the extra necessary condition for regularity $\mathcal M(0)=0$.
Using all the necessary conditions we find
\begin{eqnarray*}
I &\rightarrow& \frac{\{\mathcal M'''(0)\}^2 \xi^{*4}}{9 (1-i \xi^*)^4 (1+\xi^{*2})^2 }  \hspace{1cm}\mbox{ if } \xi^* \mbox{ finite and }\\
I &\rightarrow& 0 \hspace{4cm} \mbox{ if } \xi^* \mbox{ infinite}.
\end{eqnarray*}
So that $I$ is finite along any path reaching $r=0$.

Finally, if $\mathcal M(0)=\mathcal M'(0)=\mathcal M''(0)=0$ is satisfied then we get
\begin{eqnarray*}
K &\rightarrow& \frac{2 \{\mathcal M'''(0)\}^3 \xi^{*6}}{3 (1-i \xi^*)^2 (1+\xi^{*2})^5 } \hspace{1cm}\mbox{ if } \xi^* \mbox{ finite and }\\
K &\rightarrow& 0 \hspace{4cm} \mbox{ if } \xi^* \mbox{ infinite},
\end{eqnarray*}
what is finite along any path reaching $r=0$. $\hspace{1cm}\Box$

It is straightforward to show that the proposals for regular rotating black holes appearing in \cite{R&T}\cite{B&M}\cite{A-A}\cite{Tosh}\cite{D&G}\cite{Ghosh}\cite{LGS} do satisfy the conditions in theorem \ref{teorema}. In this way, now it can be rigourously stated that all their scalar invariants polynomial in the Riemann tensor are finite and that they do not possess scalar curvature singularities.

\section{Violations of the WEC in the core}\label{secWEC}

While in some cases a regular rotating black hole can be a solution of Einstein's equations with a non-standard energy-momentum tensor (usually with effects mainly noticeable around the core of the black hole), in most cases in the literature the RBH spacetime is not obtained from Einstein's equations. However, it is always possible to consider the existence of an \textit{effective energy-momentum tensor} defined through
\[
T_{\mu\nu}=R_{\mu\nu}-\frac{1}{2} \mathcal R g_{\mu\nu}.
\]
If we take the expression obtained for the Ricci tensor (\ref{Ricci}) one can explicit $\mathbf{T}$ for a RBH as
\[
T_{\mu\nu}=-\lambda_2 (-t_\mu t_\nu + z_\mu z_\nu)- \lambda_1 (x_\mu x_\nu+ y_\mu y_\nu).
\]
Since $\mathbf{T}$ diagonalizes in the orthonormal basis $\{\mathbf{t}, \mathbf{x}, \mathbf{y}, \mathbf{z } \}$, the RBH spacetime possesses an (effective) energy-momentum tensor of type I \cite{H&E}. The (effective) density being $\mu=\lambda_2$ and the (effective) pressures being $p_x=p_y=-\lambda_1$ and $p_z=-\lambda_2$. The weak energy conditions \cite{H&E} require $\mu\geq 0$ and $\mu+p_i\geq 0$. In other words, in this case they require
\[
\lambda_2 \geq 0 \hspace{1 cm} \mbox{and} \hspace{1 cm} \lambda_2-\lambda_1\geq 0.
\]
\begin{prop}
Assume that a regular RBH has a function $\mathcal M (r)$ that can be approximated by a Taylor polynomial around $r=0$, then the weak energy conditions should be violated around $r=0$.
\end{prop}

In order to see this,
first note that regular and admitting a Taylor polynomial means that we could write $\mathcal M (r) = \mathcal M_n r^n + O(r^{n+1}) $  with $n\geq 3$ (and $\mathcal M_n\neq 0$)\footnote{According to Taylor's theorem, the proposition requires $\mathcal M$ to be $C^n$ 
around $r=0$ and $\mathcal M^{(n+1)}$ to exist on $[0,\delta r]$ in order to be applicable \cite{B&F}.}

Second, note that the WEC imply that $\lambda_2 \geq 0$ which, using (\ref{lambda2}), is equivalent  to $\mathcal M'(r)\geq 0$. This will be satisfied around $r=0$ if $\mathcal M_n > 0$.
On the other hand, using (\ref{lambda1}) and the condition $\mathcal M_n > 0$ we get that $\lambda_2-\lambda_1\geq 0$ is equivalent, around $r=0$, to
\[
2 (r^2-a^2 \cos^2\theta) \geq (r^2+a^2 \cos^2 \theta) (n-1).
\]
This cannot be satisfied around $r=0$ for all the possible values of $\theta$. (If $n=3$ it suffices to consider $\theta\neq \pi/2$ and if $n>3$ it suffices to consider $\theta=\pi/2$). $\hspace{1cm}\Box$

\section{Conclusions}\label{secCon}

In this note we have defined a generic type of stationary rotating black hole spacetime (\ref{gIKerr}) which corresponds to the type used in most of the proposals in the literature for a (regular or not) RBH. The spacetime depends on a rotating parameter $a$ and function $\mathcal M(r)$, whose specific form differentiates the different proposals. We have shown that the spacetime for a RBH is Petrov type D and Segre Type [(1,1) (1 1)]. As a consequence, we have concluded that there are only 6 real second order invariants algebraically independent for these spacetimes. Therefore, in order to guarantee that a RBH spacetime does not possess a scalar curvature singularity none of these invariants has to diverge along any path in the spacetime. We have seen that the 6 invariants can be collected into two real invariants $\mathcal R$ and $I_6$ and two complex invariants $I$ and $K$. We have shown that the only possible source of singularities is the \textit{ring} defined in Boyer-Lindquist-like coordinates by $(r=0,\theta=\pi/2)$, so that we have only needed to analyze the behaviour of the invariants as the ring was approached along different paths. This has led us to a simple necessary and sufficient condition on $\mathcal M(r)$ and its derivatives in order to avoid scalar curvature singularities: $\mathcal M(0)=\mathcal M'(0)=\mathcal M''(0)=0$.

The avoidance of singularities in the RBH spacetimes satisfying these conditions suggested that some of the requirements in the standard singularity theorems are violated. In effect, we have been able to show that, if a regular black hole has a function $\mathcal M$ admitting a Taylor polynomial around $r=0$, then the weak energy conditions should be violated in the core of the black hole.

\section*{Acknowledgements}
The authors acknowledge the financial support of the Ministerio de Econom\'{\i}a y Competitividad (Spain), projects MTM2014-54855-P.


\end{document}